\documentclass[twocolumn]{aastex631}
\usepackage{graphicx}	
\usepackage{amsmath}	
\usepackage[T1]{fontenc}
\usepackage{newtxtext,newtxmath}

\usepackage{amssymb}	
\usepackage{gensymb}
\usepackage{color}
\usepackage{natbib}
\usepackage{enumitem}
\usepackage{mathtools}
\usepackage{threeparttable}
\usepackage{textcomp,booktabs}   
\usepackage{multirow}
\usepackage{verbatim}
\usepackage{CJK}

\usepackage[mathlines]{lineno} 

\begin{document}
\begin{CJK*}{UTF8}{gkai}

\title{Broadband Properties of the Harmonic of Type-C Quasi-periodic Oscillation in MAXI J1348$-$630}

\correspondingauthor{Xin-Lei Wang \& Zhen Yan \& Ren-Yi Ma}
\email{wxl2@jgsu.edu.cn, zyan@shao.ac.cn, ryma@xmu.edu.cn}
\author{Xin-Lei Wang (王鑫磊)}
\affil{School of Mathematics and Physics, Jinggangshan University, Ji'an, Jiangxi 343009, China}

\author{Zhen Yan (闫震)}
\affil{Shanghai Astronomical Observatory, Chinese Academy of Sciences, 80 Nandan Road, Shanghai 200030, China}
\affil{SHAO-XMU Joint Center for Astrophysics,  Xiamen, Fujian 361005, China}

\author{Jun-Feng Wang (王俊峰)}
\affil{Department of Astronomy and Institute of Theoretical Physics and Astrophysics, Xiamen University, Xiamen, Fujian 361005, China}
\affil{SHAO-XMU Joint Center for Astrophysics,  Xiamen, Fujian 361005, China}

\author{Fu-Guo Xie (谢富国)}
\affil{Shanghai Astronomical Observatory, Chinese Academy of Sciences, 80 Nandan Road, Shanghai 200030, China}
\affil{SHAO-XMU Joint Center for Astrophysics,  Xiamen, Fujian 361005, China}

\author{Zhan-Yi Liu (刘展翼)}
\affil{Department of Astronomy and Institute of Theoretical Physics and Astrophysics, Xiamen University, Xiamen, Fujian 361005, China}
\affil{SHAO-XMU Joint Center for Astrophysics,  Xiamen, Fujian 361005, China}

\author{Ren-Yi Ma (马任意)}
\affil{Department of Astronomy and Institute of Theoretical Physics and Astrophysics, Xiamen University, Xiamen, Fujian 361005, China}
\affil{SHAO-XMU Joint Center for Astrophysics,  Xiamen, Fujian 361005, China}

\begin{abstract}

Harmonics are common features of quasi-periodic oscillations (QPOs) in black hole X-ray binaries; however, their physical origins remain poorly understood. Using broadband \textit{Insight}-HXMT data, we investigated the Type-C QPO harmonic in MAXI J1348$-$630.
The harmonic is significantly detected exclusively during the hard intermediate state and displays prominent energy-dependent properties: while it is strong in the soft X-ray band ($<$10 keV) with its fractional rms amplitude even exceeding that of the fundamental QPO, it is much weaker in the hard X-ray band ($>$10 keV), where its rms amplitude is generally several times lower than that of the fundamental.
Furthermore, the harmonic shows no significant phase coupling with the fundamental in the soft X-ray band, whereas strong coupling is present in the hard X-ray band.
These features point to a complex, energy-dependent origin for the harmonic.
We propose that the hard X-ray harmonic may arise from nonlinear distortion of the fundamental waveform within the corona, while the soft X-ray harmonic is likely produced via a distinct physical process, such as the reflection emission from the inner disk.
\end{abstract}



\section{Introduction}
\label{sec:intro}
\end{CJK*}

Quasi-periodic oscillations (QPOs) are among the most representative
timing phenomena observed in black hole X-ray binaries (BHXRBs).
They manifest as narrow peaks in the power density spectrum (PDS) of the X-ray light curve, revealing quasi-periodic modulations in the X-ray emission on short timescales.
Based on their centroid frequencies, QPOs are classified into low-frequency QPOs (LFQPOs; a few mHz to tens of Hz) and high-frequency QPOs (HFQPOs; tens to hundreds of Hz) \citep{psaltis_1999, Remillard99a}.

LFQPOs are commonly divided into Types-A, -B, and -C based on their centroid frequency, quality factor \(Q\) (defined as the ratio of the centroid frequency to the full width at half maximum), fractional root-mean-square (rms) amplitude, and accompanying background noise \citep{Casella04, Casella05, Motta_2011}. 
Among them, Type-C QPOs are the most frequently detected during the hard state (HS) and hard intermediate state (HIMS), appearing as relatively narrow peaks with strong rms amplitudes and often accompanied by prominent band-limited noise (BLN) \citep{Belloni02a, Remillard_2006, Ingram19}. 
Type-B QPOs are typically found in the soft intermediate state (SIMS) and also show strong rms amplitudes and narrow peaks \citep[e.g.,][]{Casella05, Stevens_2016}; recent observations have reported their sporadically simultaneous presence with Type-C QPOs \citep{Pei_2025_simul, wang_2026_co}. 
Type-A QPOs are the least frequently detected among the three types of LFQPOs, appearing in the SIMS and soft state (SS) as weak, broad peaks associated with red noise components \citep{Homan01, Casella04, zhang_2023}.

A common observational feature of QPOs is the presence of harmonics, which appear as additional peaks in the PDS at integer multiples of the fundamental frequency (e.g., $2\nu_{\mathrm{QPO}}$, $3\nu_{\mathrm{QPO}}$), as well as subharmonics at fractional multiples (e.g., $\nu_{\mathrm{QPO}}/2$). 
Harmonic peaks are usually labeled according to their frequency ratios relative to the fundamental frequency. 
In this work, we adopt the convention of \citet{Casella04} and \citet{Ingram19}, where the $n$th harmonic corresponds to a frequency of $n\nu_{\mathrm{QPO}}$.
Under this convention, the peak at $2\nu_{\mathrm{QPO}}$ is the second harmonic. For convenience, hereafter 
the unqualified term "harmonic" refers specifically to the second harmonic, since it is the most commonly detected harmonic.

Theoretically, there are two main possible origins for the harmonic: either it is driven by a physical process with a period half that of the fundamental \citep[e.g.,][]{Axelsson16, Doesburgh_2020}, or it arises simply from the nonlinear distortion of the fundamental waveform, which mathematically generates a harmonic peak in Fourier space \citep[e.g.,][]{Shui_2021}. In the latter scenario, the harmonic is naturally expected to exhibit observational properties similar to those of the fundamental (such as energy dependence), and a strong and stable phase coupling between the two components is also expected.

Previous studies have analyzed this phase coupling across several sources using different methods. Using short-segment Fourier transform and statistical analysis, \citet{Ingram_2014} found that the phase difference between the harmonic and fundamental in GRS 1915+105 clusters around a preferred value, indicating persistent phase coupling. Similar results were also obtained using a mode-decomposition method applied to MAXI J1820+070 \citep{Shui_2023_hht}. Furthermore, several studies utilizing bispectral analysis (detailed in \autoref{sec:data}) have reported strong phase coupling between the fundamental and its harmonic \citep{Maccarone_2011, Arur_2019, zhuhaifan_2024_2}. These results suggest that the harmonic may share a common origin with the fundamental, such as through nonlinear waveform distortions.

However, substantial observational evidence has shown that the harmonic often exhibits properties distinct from those of the fundamental, complicating its original interpretations. For example, the harmonic of Type-C QPO is often weakened or even undetectable in both the softest (e.g., $\lesssim 3$ keV) and hardest (e.g., $\gtrsim 10$ keV) energy bands, in contrast to the fundamental, which is detected over a broader energy range \citep{Rodriguez_2002_1915, Jithesh_2019, Debnath_2024, zhuhaifan_2024_1}.
Frequency-resolved spectral analyses further show that the harmonic spectrum is occasionally softer than that of the fundamental QPO but harder than the thermal disk emission, pointing to a relatively cool Comptonizing region \citep{Axelsson14, Axelsson16}.
In addition, the harmonic often exhibits a more stable and larger hard phase lag compared to the fundamental \citep{Casella04, zhangliang_2020, Debnath_2024}. These distinct features indicate that the harmonic originates from a more complex mechanism, and further investigations into the harmonic are necessary to fully unravel its physical origin.

\section{MAXI J1348$-$630}
\label{MAXIJ1348}
\begin{figure*}
\centering
\includegraphics[width=0.95\linewidth]{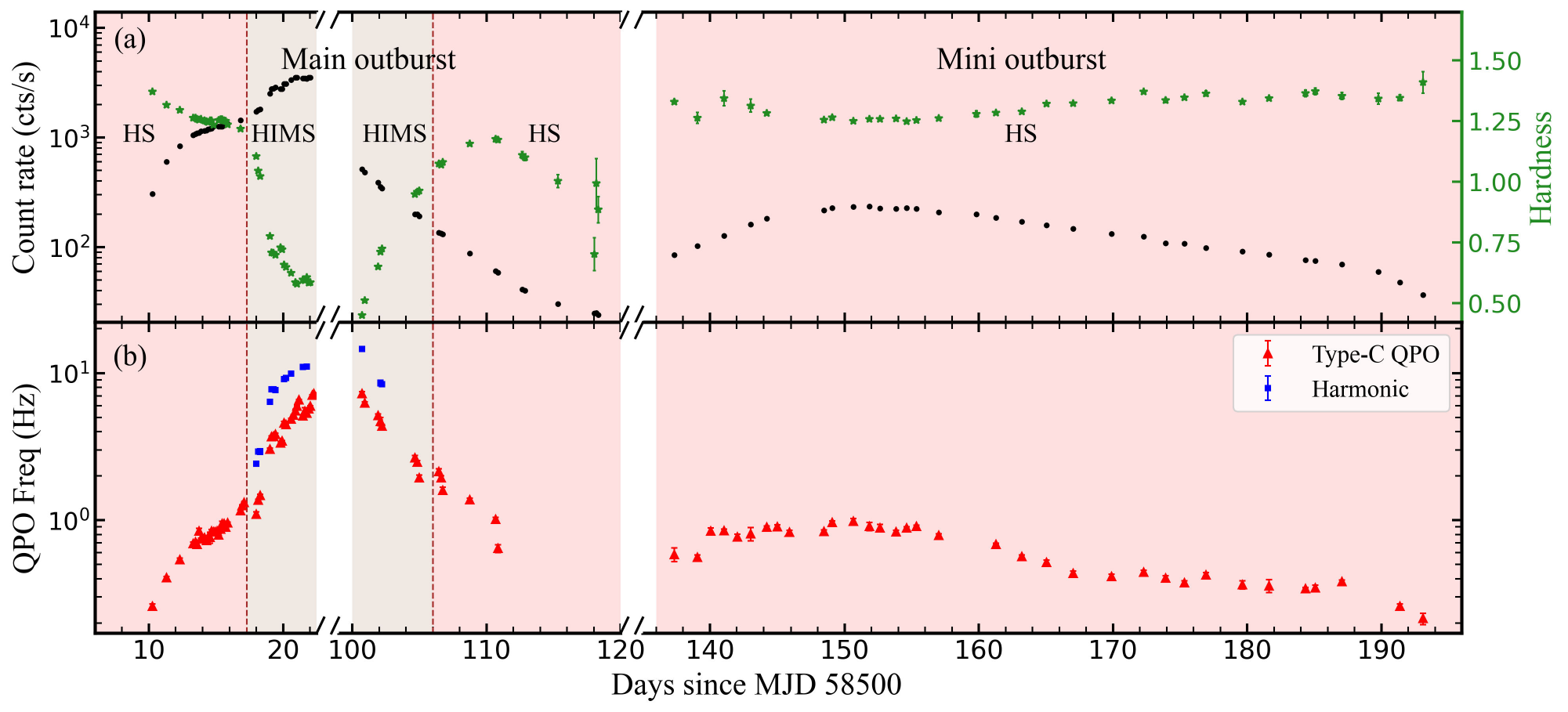}
\caption{(a) Evolution of the 2019 outburst of MAXI J1348$-$630. The black dots and green stars represent the LE count rate and hardness, respectively, where the hardness is defined as the ratio of the count rates in the 3.5--10 keV and 2--3.5 keV bands. (b) Evolution of the frequencies of the Type-C QPO and its harmonic. Only harmonics with a detection significance of \(\ge 3\sigma\) are plotted.}
\label{fig:outburst-evolution}
\end{figure*}

MAXI J1348$-$630 was discovered during its 2019 outburst \citep{Yatabe2019} and was subsequently identified as a black hole X-ray binary \citep{Zhang2020}. 
As shown in \autoref{fig:outburst-evolution}, the source remained in the HS from MJD 58510.3 until 58517.2, after which it evolved into the HIMS and persisted until MJD 58522.3. Subsequently, the outburst evolved through the SIMS and SS, re-entered the HIMS after MJD $\sim$58600, and returned to the HS after MJD $\sim$58606 \citep{Zhang2020, You_2024, Carotenuto_2025}. 
Through modeling the jet kinematics, its inclination angle was constrained to $\sim$\(29.3^\circ\) \citep{Carotenuto2022MNRAS}.
The source exhibited a $\sim$110-day main outburst followed by several mini-outbursts, during which all three types of LFQPOs were detected \citep{Zhang2020}.
Previous studies have shown that Type-B and Type-A QPOs in this source are related to Comptonization processes, possibly associated with a jet \citep{zhang2021,Garc2021,Bellavita_2022,Liu_2022} or an extended corona \citep{zhang_2023}, respectively.
For Type-C QPOs, they were detected from the onset of the HS and exhibit a clear frequency evolution from $\sim$0.3 Hz to about 7 Hz toward the end of the HIMS \citep{Jithesh2021,Alabarta_2022,wang_2026_CQPO}.
Their fractional rms spectra generally rise toward higher energies, with measured hard phase lags consistently pointing to an origin in a horizontally extended Comptonizing region \citep{Alabarta_2022,Alabarta24}.

Although previous studies have established the comprehensive timing and spectral properties of Type-C QPOs, their harmonic features have not yet been systematically investigated.
\textit{Insight}-HXMT performed extensive monitoring of MAXI J1348$-$630 and provided abundant broadband data covering 1--250 keV, enabling detailed analysis of the harmonic in this source.
In the following, we describe the data reduction and analysis methods in \autoref{sec:data}, present the analysis results in \autoref{sec:results}, discuss their implications for the physical origin of the harmonic in \autoref{sec:discussion}, and conclude our work in \autoref{sec:conclusion}. 

\section{DATA REDUCTION and analysis}
\label{sec:data}
\textit{Insight}-HXMT is a satellite that provides simultaneous broadband coverage from 1 to 250 keV via its three instruments: the Low Energy X-ray Telescope (LE), Medium Energy X-ray Telescope (ME), and High Energy X-ray Telescope (HE). These three telescopes have time resolutions of 1 ms, 240 $\mu$s, and 4 $\mu$s, and effective areas of 384, 952, and 5100 cm$^2$, respectively \citep{zhang_overview_2020}. Each observation from \textit{Insight}-HXMT consists of several data segments, referred to as "exposures'', which are identified by unique IDs.

For this study, we use the following energy bands for the three telescopes: 1--10 keV (LE), 10--30 keV (ME), and 27--200 keV (HE). Hereafter, we use the abbreviations LE, ME, and HE exclusively to refer to these corresponding energy bands.
We processed the raw data using the \textit{Insight}-HXMT Data Analysis Software package (\texttt{HXMTDAS}) version 2.05. Standard procedures were applied for calibration and filtering. Tools within this software package (e.g., \texttt{helcgen, hebkgmap}) were used to generate the raw light curves and their corresponding background light curves with a time resolution of 1/256 seconds.

We generated PDSs from the raw light curves using the \texttt{powspec} tool with different segment durations. For the HS, we used a segment duration of 64 s, corresponding to a frequency resolution of $\sim$0.016 Hz. For the HIMS, since the type-C QPOs exhibit higher frequencies (above 1 Hz), we adopted a shorter segment duration of 16 s to increase the number of segments for averaging, thus improving the signal-to-noise ratio (S/N) of the PDSs. The resulting PDSs were averaged and logarithmically rebinned with a geometric factor of 1.03.
All PDSs were normalized according to the Leahy normalization \citep{Leahy1983}, and the Poisson noise was subtracted. To enable spectral fitting within \texttt{XSPEC} (version 12.10.1), the PDSs were converted into a compatible format using the \texttt{flx2xsp} tool \citep{Ingram12}.

\begin{figure*}
\centering
\includegraphics[width=0.98\linewidth]{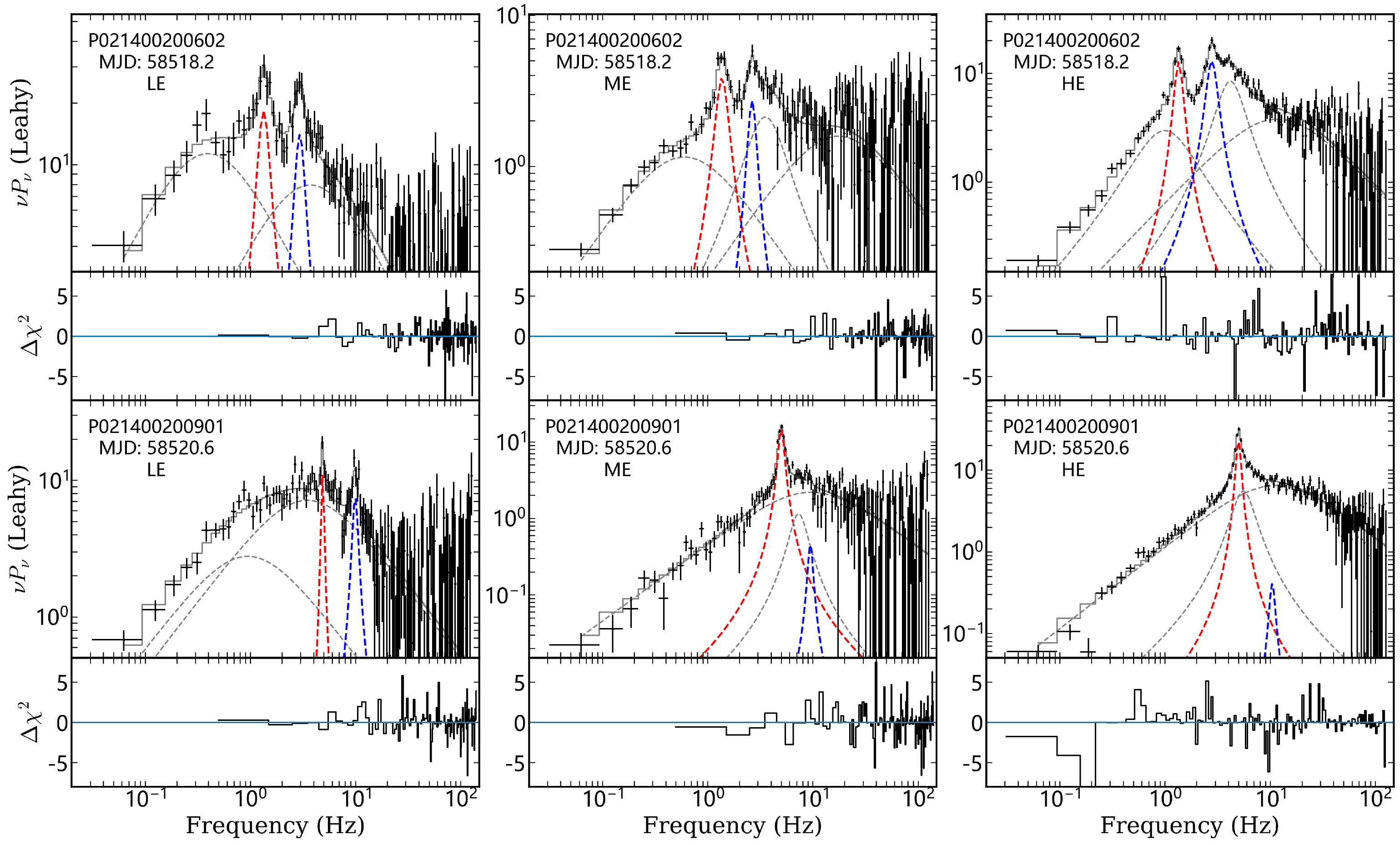}
\caption{PDSs for two exposures, P021400200602 (upper) and P021400200901 (lower), across the LE, ME, and HE bands. The Type-C QPO and its harmonic are indicated by red and blue dashed lines, respectively. From the upper to the lower panel, the harmonic shows a dramatic change in energy dependence: it is strong across all bands in the earlier exposure (P021400200602) but has nearly vanished in the ME and HE bands in the later one (P021400200901).}
\label{fig:pds}
\end{figure*}

The PDS were fitted with a multiple \texttt{lorentz} model in \texttt{XSPEC}, and the best-fitting parameters and their 1$\sigma$ uncertainties are retrieved from the chains generated by the Markov Chain Monte Carlo (MCMC) algorithm \texttt{emcee} implemented in \texttt{XSPEC}\footnote{\url{https://github.com/zoghbi-a/xspec_emcee}}. The fractional rms amplitude was calculated as $\sqrt{\texttt{norm}/\langle C_{\rm raw} \rangle}$ 
\citep{vanderKlis1988,VanDerKlis89}, 
where $\texttt{norm}$ is the integrated power of the \texttt{lorentz} component in Leahy normalization, and $\langle C_{\rm raw} \rangle$ is the mean count rate of the raw light curve. 
To account for background dilution, the resulting rms values were corrected by a factor of $(S+B)/S$, where $S$ and $B$ denote the source and background count rates, respectively \citep{Belloni1990}.
The significance of QPOs was determined by taking the ratio of the integrated power from the \texttt{lorentz} component (i.e., the \texttt{norm} parameter) to its negative $1\sigma$ uncertainty, following the approach adopted in previous works \citep{Motta_2015,zhangliang_2020,wangjingyi_2024,Kumar_2024}.

To quantify the nonlinear phase coupling between the fundamental and harmonic components, we employ bispectral analysis \citep{Maccarone_2002}.
For a light curve \(x(t)\) divided into \(K\) segments, the bispectrum is defined as
\begin{equation}
B(\nu_1,\nu_2) = \frac{1}{K} \sum_{i=0}^{K-1} X_i(\nu_1) X_i(\nu_2) X_i^*(\nu_1+\nu_2) ,
\end{equation}
where \(X_i(\nu)\) is the Fourier transform of the \(i\)-th segment at frequency \(\nu\), and \(X_i^*(\nu)\) denotes its complex conjugate.
This quantity characterizes the phase coupling among the three frequency components $\nu_1$, $\nu_2$, and $\nu_1+\nu_2$.
Since the bispectrum depends on the signal amplitude and noise level, it is commonly normalized to yield the bicoherence:
\begin{equation}
b^2(\nu_1,\nu_2) =
\frac{\left| \sum X_i(\nu_1) X_i(\nu_2) X_i^*(\nu_1+\nu_2) \right|^2}
{\sum |X_i(\nu_1) X_i(\nu_2)|^2 \sum |X_i(\nu_1+\nu_2)|^2} \,.
\end{equation}
The bicoherence takes values between 0 and 1, with larger values indicating stronger nonlinear phase coupling.
For the QPO harmonic analysis, we focus on the configuration $\nu_1=\nu_{\rm QPO}$, $\nu_2=\nu_{\rm QPO}$, and $\nu_1+\nu_2=2\nu_{\rm QPO}$, which allows us to directly measure the phase relation between the fundamental and its harmonic.

The bispectral analysis was performed using the Python package \texttt{higher-spectrum}\footnote{\url{https://github.com/synergetics/spectrum}}. To optimize the frequency resolution and maintain a high S/N, the segment length was adjusted (e.g., to 16 s) according to the QPO frequency during different spectral states. Furthermore, a sliding window with a 70\% overlap between consecutive segments was employed to reduce the variance of the bispectral estimates. In addition to the bispectral analysis, we investigated the phase lag using the \texttt{AveragedCrossspectrum} class from the \texttt{Stingray} package \citep{Huppenkothen_2019}. The segment lengths for computing the cross spectra were similarly set to either 64 s or 16 s, depending on the respective QPO frequency. Finally, the resulting cross spectra were logarithmically rebinned with a geometric factor of 1.03.

\section{results}
\label{sec:results}
While \autoref{fig:outburst-evolution}(a) shows the overall outburst evolution of MAXI J1348$-$630, panel (b) displays the corresponding frequency evolution of the Type-C QPO (red points). The QPO frequency increased from about 0.24 Hz at the beginning of the HS to about 7 Hz by the end of the rising HIMS, and subsequently decreased to \(\sim\)0.7 Hz during the decay phase of the main outburst \citep{Alabarta_2022, wang_2026_co}.

\subsection{State-Dependence of the Harmonic}
\label{sec:State dependence}
During the initial HS, no significant harmonic of the Type-C QPO was detected, with no obvious peak appearing at twice the QPO frequency in the PDSs. 
To quantitatively evaluate the significance and strength of the harmonic during this phase, we included an additional \texttt{lorentz} component in the fitting model, with its centroid frequency and FWHM fixed at twice those of the Type-C QPO. 
The significance of this component consistently remains below \(2\sigma\) in all three energy bands, with its fractional rms amplitudes typically around 1--1.5\% and never exceeding 2.5\%, confirming that the harmonic in this state is weak and of low significance.

\begin{figure*}
\centering
\includegraphics[width=1\linewidth]{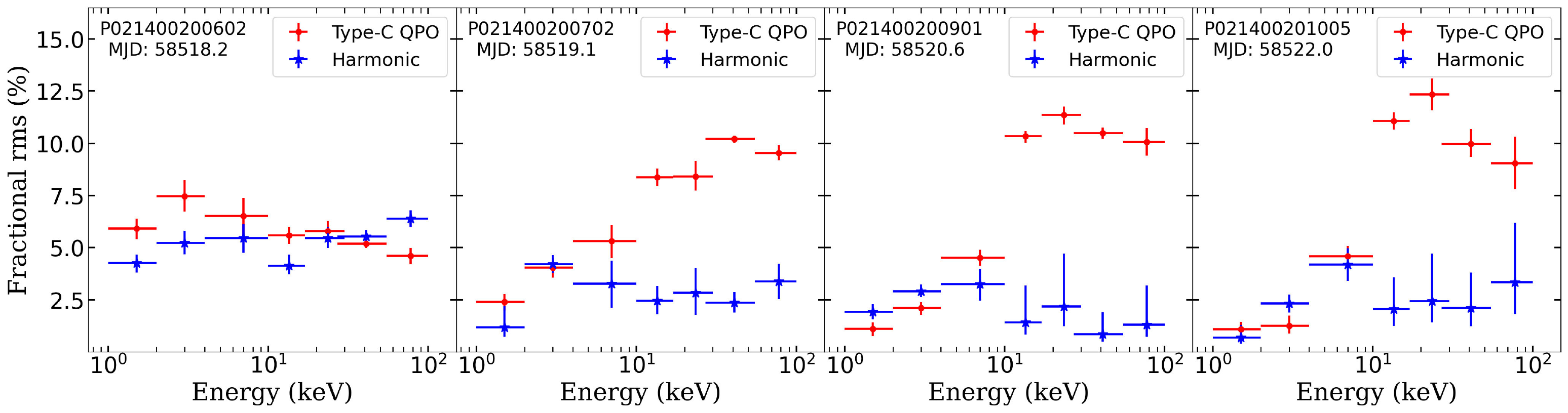}
\caption{Evolution of the fractional rms spectra for the Type-C QPO and its harmonic. The leftmost panel corresponds to the initial emergence of the harmonic, where it exhibits comparable rms amplitude across energy bands. In subsequent panels, the harmonic becomes weaker at both the softest and hard bands.}
\label{fig:rms-spectra}
\end{figure*}

After MJD 58517.2, the rapid drop in the hardness signifies the onset of the transition to the HIMS. 
In the early phase of the HIMS around MJD 58518, \textit{Insight}-HXMT performed an observation (P0214002006) consisting of three exposures. 
A clear harmonic is detected in the PDSs of all three exposures. 
As an illustration, the PDSs of P021400200602 are shown in the upper row of \autoref{fig:pds}. 
In addition to the Type-C QPO, a strong harmonic is present across the broad energy band from LE to HE, with detection significances of $4.2\sigma$, $2.7\sigma$, and $7.0\sigma$, respectively, and fractional rms amplitudes exceeding 4.5\% in all bands.

Subsequently, the harmonic is frequently detected in the HIMS during the rising phase, as indicated by the blue points in \autoref{fig:outburst-evolution}(b) (see \autoref{tab:harmonic_fits} for detailed fitting parameters).
We show the PDS of exposure P021400200901 (MJD 58520.6) in the lower row of \autoref{fig:pds} as another example.
In the LE band of this exposure, the harmonic is detected at a significance of $4.9\sigma$. 
It should be noted that the harmonic is much weaker in the ME and HE bands at this epoch, possibly indicating an evolution in its energy dependence.  
Details of this behavior are presented in \autoref{sec:Energy dependence}.

A similar pattern is also observed during the decay phase of the main outburst and in the mini outburst. 
During the decay-phase HIMS, the harmonic is detected in some exposures at a significance above $3\sigma$, despite the relatively low photon count rate. 
In contrast, no significant harmonic is detected in the HS (all below $2\sigma$). 
These results demonstrate a clear spectral state dependence of the harmonic in MAXI J1348$-$630: it remains weak and insignificant in the HS, but becomes stronger and highly significant in the HIMS.

\subsection{Energy-Dependent Evolution of the Harmonic}
\label{sec:Energy dependence}
As shown in the PDSs of P021400200602 in \autoref{fig:pds}, the harmonic is strong across all three energy bands at the onset of the HIMS.
The corresponding rms values in the LE, ME, and HE bands are $5.74^{+0.89}_{-0.68}\%$, $4.59^{+1.03}_{-0.86}\%$, and $4.64^{+0.38}_{-0.33}\%$, respectively.
These values are comparable to those of the fundamental, which are $7.01^{+0.68}_{-0.77}\%$, $6.30^{+0.45}_{-0.41}\%$, and $5.03^{+0.34}_{-0.34}\%$ in the same bands.
This behavior differs from the typical case, where the harmonic is several times weaker than the fundamental \citep{Ingram19,Doesburgh_2020}.

About one day after the early HIMS observation P0214002006, around MJD 58519, the hardness of the source decreases from $\sim$1.1 to $\sim$0.7 (see \autoref{fig:outburst-evolution}).
After this stage, the QPO properties change substantially, as illustrated by the example exposure P021400200901 in the lower row of \autoref{fig:pds}.
The Type-C QPO becomes prominent in the hard X-ray bands (ME and HE) but is relatively weak in the LE band.
The harmonic shows the opposite behavior, being strongly suppressed in the hard X-ray bands while remaining relatively strong in the LE band.
Specifically, the rms values of the fundamental in P021400200901 are $2.21^{+0.18}_{-0.20}\%$, $10.83^{+0.30}_{-0.33}\%$, and $10.54^{+0.64}_{-0.55}\%$ in the LE, ME, and HE bands, respectively, while those of the harmonic are $2.33^{+0.32}_{-0.24}\%$, $1.36^{+1.17}_{-0.51}\%$, and $1.14^{+1.07}_{-0.49}\%$.

To further investigate this behavior, we examine the fractional rms spectrum in more detailed energy bands.
The energy bands are divided as follows: 1--2 keV, 2--4 keV, and 4--10 keV for LE; 10--17 keV and 17--30 keV for ME; and 27--55 keV and 55--100 keV for HE.
For energy bands where the harmonic is too weak to be well constrained, we estimate its strength by fixing its centroid frequency and FWHM to twice those of the fundamental.

\begin{figure*}
\centering
\includegraphics[width=1\linewidth]{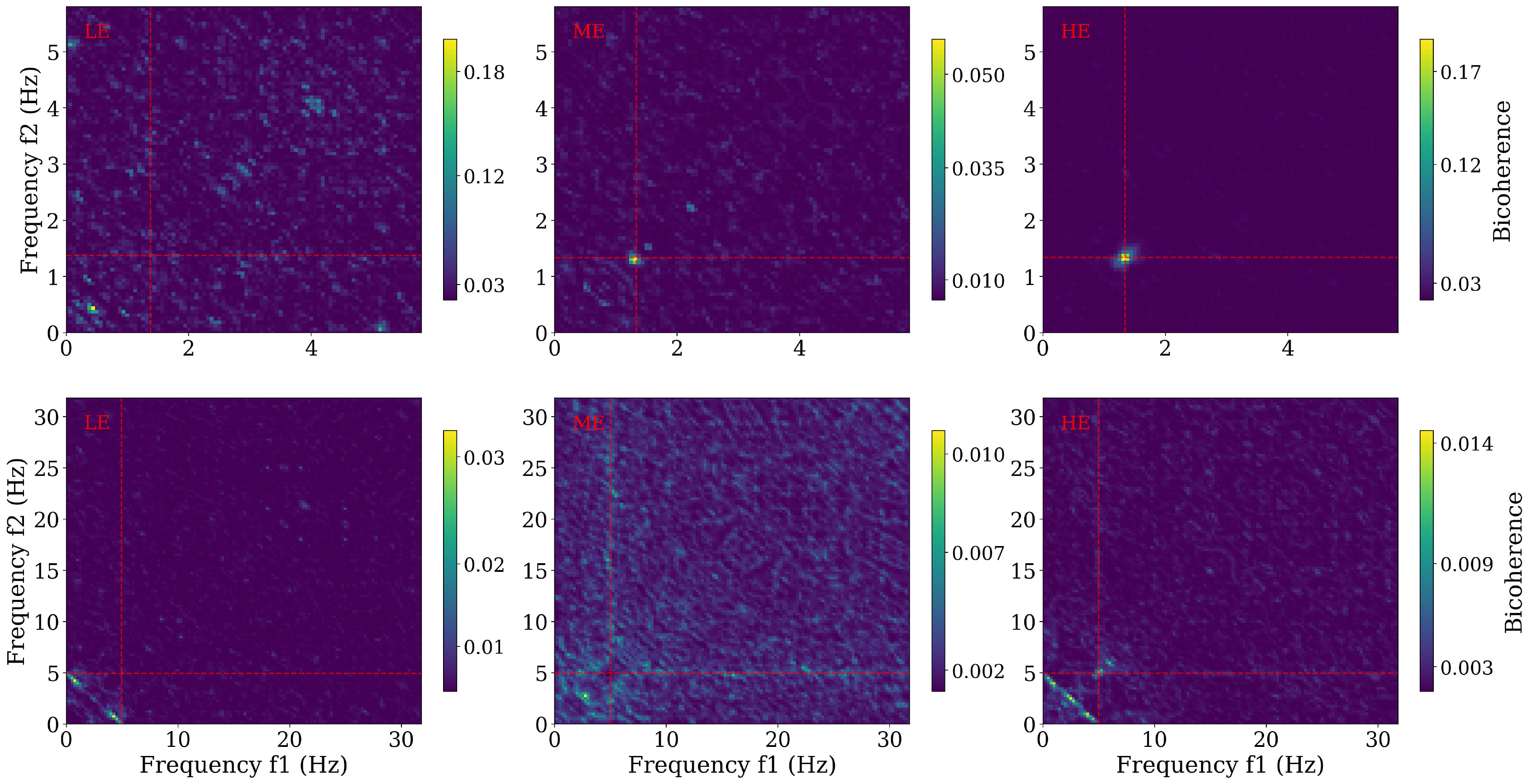}
\caption{Bispectral analysis across three energy bands for two exposures (P021400200602, top; P021400200901, bottom). Their corresponding PDS are shown in \autoref{fig:pds}. The red dashed lines mark the centroid frequency of the Type-C QPO. No significant bicoherence is detected in the LE band across both exposures. In contrast, strong bicoherence persists in the HE band, even when the harmonic is barely visible in the HE PDS for the later exposure.}
\label{fig:bicoherence}
\end{figure*}

\begin{figure}
\centering
\includegraphics[width=0.8\linewidth]{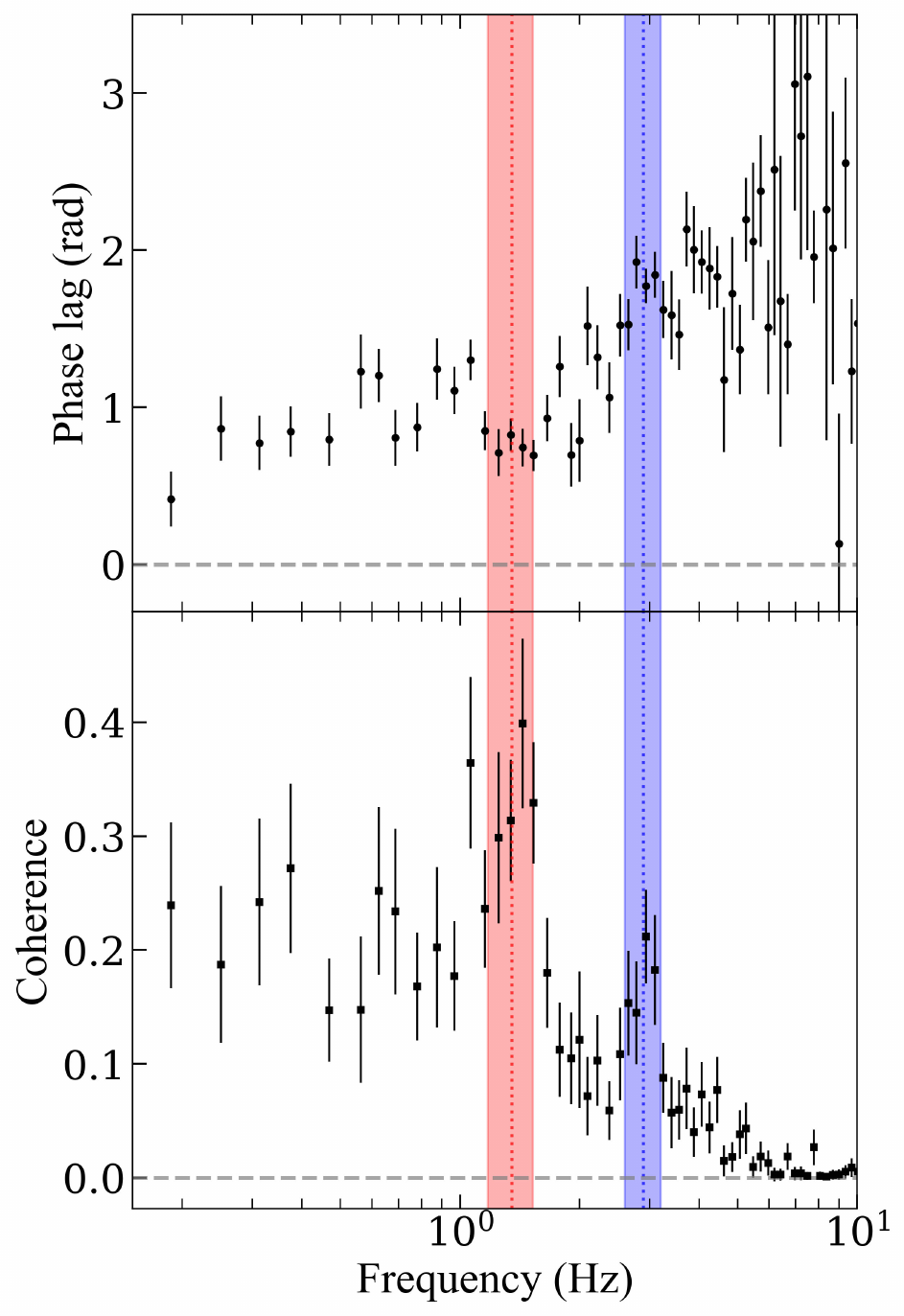}
\caption{Phase lags and coherence of the type-C QPO and its harmonic. Top: Phase lags (27--55 keV relative to 1--2 keV) as a function of frequency for exposures P021400200602 and P021400200603. The frequency ranges of the fundamental QPO and its harmonic are highlighted in red and blue shading, respectively. Bottom: Coherence coefficient versus frequency for the same energy bands and exposures.}
\label{fig:phase-lag}
\end{figure}

\autoref{fig:rms-spectra} presents the fractional rms spectra of the Type-C QPO and its harmonic for four representative exposures spanning from the early to the late HIMS during the rising phase. 
In the first panel (P021400200602), the harmonic rms amplitude remains approximately constant at $\sim$5\% across all energy bands. 
In the subsequent three panels, however, the harmonic rms amplitude is strongly suppressed in the hard X-ray bands (10 keV and above), with most values falling below 3\%. 
In the soft X-ray bands, the rms amplitude is also reduced in the softest 1--2 keV band, with the peak generally located at 4--10 keV. 
This indicates that, except for the onset of the HIMS, the harmonic rms spectrum peaks at intermediate energies rather than at either the softest or hardest bands.

Taken together, these results reveal a rapid evolution in the energy dependence of the harmonic during the HIMS, from being strong across all bands to being prominent at intermediate X-ray energies of 4--10 keV.

\subsection{Bispectral Analysis of the Harmonic}
\label{sec:bicoherence}
Given the pronounced state- and energy-dependent behavior of the harmonic, we further investigate its origin by examining the phase relationship between the harmonic and the fundamental using bispectral analysis.
As shown in \autoref{fig:bicoherence}, we again analyze the two representative exposures P021400200602 from the early HIMS and P021400200901 from the later HIMS, for which the harmonic rms spectra differ significantly between these two stages (see \autoref{sec:State dependence} and \autoref{sec:Energy dependence}).
Due to the distinct evolutionary behavior of the harmonic across different energy bands, we perform the phase coupling analysis separately for the LE, ME, and HE bands.

The bicoherence results for these two exposures are presented in \autoref{fig:bicoherence}.
To highlight the phase coupling between the fundamental and the harmonic, we mark the fundamental frequency with red dashed lines; the bicoherence amplitude at their intersection corresponds to $(\nu_{\mathrm{QPO}}, \nu_{\mathrm{QPO}})$, thereby quantifying the phase coupling between the fundamental ($\nu_{\mathrm{QPO}}$) and its harmonic ($2\nu_{\mathrm{QPO}}$, arising from $\nu_{\mathrm{QPO}}+\nu_{\mathrm{QPO}}$).

Despite the clear differences in the PDSs, the bispectral results of the two exposures are broadly similar.
In the LE band, no significant bicoherence signal is detected at $(\nu_{\mathrm{QPO}}, \nu_{\mathrm{QPO}})$ in either exposure, indicating little or no nonlinear phase coupling between the fundamental and the harmonic.
In contrast, for the ME and HE bands of P021400200602, where strong harmonics are observed in the PDSs (\autoref{fig:pds}), the bispectral analysis reveals strong phase coupling between the harmonic and the fundamental.
For P021400200901, although the harmonic becomes very weak in the PDSs of the ME and HE bands, a relatively strong bicoherence signal remains at $(\nu_{\mathrm{QPO}}, \nu_{\mathrm{QPO}})$ in the HE band, indicating that the phase coupling persists even when the harmonic is barely detectable in the PDS.

Overall, in contrast to the pronounced evolution of the harmonic fractional rms spectrum from the early to the late HIMS, the phase coupling between the harmonic and the fundamental remains remarkably stable: it is consistently absent in the LE band and strong in the HE band, while the ME band shows intermediate behavior, likely reflecting its transitional energy range.

\subsection{Phase Lag of the Harmonic}
\label{sec:lag}
We further investigate the phase lag of the harmonic and compare it with that of the fundamental.
We perform cross-spectral analysis using two well-separated energy bands, 1--2 keV and 27--55 keV, for comparison.
Since the harmonic becomes extremely weak in the hard X-ray bands during the later HIMS (see \autoref{fig:rms-spectra}), the coherence at the harmonic frequency is too low to yield reliable phase lag measurements.
We therefore use the exposure P021400200602, as both the fundamental and the harmonic are significantly detected in the soft and hard X-ray bands, ensuring high coherence and reliable phase lag estimates.
Furthermore, given the similar QPO frequencies in the subsequent exposure P021400200603, we combine the two to improve the S/N.

The results are shown in \autoref{fig:phase-lag}, where the upper panel presents the phase lag spectrum and the lower panel shows the coherence spectrum.
The harmonic (blue shaded region) exhibits a larger hard lag than the fundamental (red shaded region), consistent with results reported for other sources \citep[e.g.,][]{Debnath_2024}.
We further attempt to derive the phase lag values for the fundamental and harmonic. Following \citet{Mendez24}, the contributions from the broadband noise should be subtracted from the real and imaginary parts rather than directly in phase space. We therefore apply a localized baseline subtraction to the real and imaginary components separately as an approximation to calculate the phase lags.
Specifically, we define the QPO characteristic frequency range as $\nu_{\rm QPO} \pm \tfrac{1}{2}\,\text{FWHM}$, and adopt the average values over one FWHM on either side of this interval as the noise baseline in the real and imaginary parts. We then calculate the intrinsic QPO phase lags from these background-subtracted real and imaginary parts.
The phase lag of the fundamental is $0.43\pm0.29$ rad, corresponding to a time lag of $0.05\pm0.03$ s, while the phase lag of the harmonic is $2.16\pm0.19$ rad, corresponding to a time lag of $0.12\pm0.01$ s.

\section{Discussion}
\label{sec:discussion}
The main properties of the harmonic of the Type-C QPO in MAXI J1348$-$630 can be summarized as follows:
\begin{itemize}
    \item The harmonic is not significantly detected during the HS; however, as the source transitions into the HIMS, it becomes persistently detected, with its fractional rms occasionally exceeding that of the fundamental.
    \item Except for the onset of the HIMS, the fractional rms spectrum of the harmonic is much softer than that of the fundamental, peaking at 4--10 keV.
    \item The harmonic and the fundamental show no significant phase coupling in the soft X-ray band, but exhibit strong coupling in the hard X-ray band.
    \item The harmonic exhibits a larger hard phase lag compared to the fundamental.
\end{itemize}

\subsection{Universality and Diversity of the Harmonic Across Different BHXRBs}
\label{sec:discussion-comparison}
Before investigating the specific formation mechanism of the harmonic in MAXI J1348$-$630, we first compare its properties with those reported in other BHXRBs.
The harmonic behavior observed in MAXI J1348$-$630 is similarly present in GX 339$-$4. During the 2007 and 2010 outbursts of GX 339$-$4, \citet{Shui_2021} reported that the harmonic of the Type-C QPO was nearly absent in the HS but emerged prominently upon the source entering the HIMS, with its fractional rms amplitude exceeding that of the fundamental QPO. Furthermore, \citet{Axelsson16} reported that during the early HIMS of the 2007 outburst, the harmonic was strong across soft (3--5 keV) and hard (10--30 keV) X-ray bands; as the outburst evolved, it rapidly weakened in the hard band while remaining prominent in the soft band. These state-dependent and energy-dependent evolutionary features are highly consistent with our findings in MAXI J1348$-$630.
A similar state-dependent harmonic behavior has also been found in MAXI J1535$-$571 \citep{Mereminskiy_2018}, suggesting that the physical process underlying the harmonic formation may be universal among these BHXRBs.

However, the above properties of the harmonic are not uniformly shared across all BHXRBs. In contrast, for a number of other sources, the harmonic is significantly detected at the beginning of the HS, almost always accompanying the Type-C QPO, and its amplitude remains much weaker than that of the fundamental. This distinct behavior has been observed in sources such as MAXI J1820+070 \citep{Ma_2023}, Swift J1727.8$-$1613 \citep{Yu_2024}, XTE J1859+226 \citep{Casella04}, and XTE J1550$-$564 \citep{Rao_2010}. Such diverse manifestations suggest that the harmonic might have different formation mechanisms. For example, one possible factor is the geometric effect arising from the source inclination angle, as the projected area of a precessing accretion flow naturally produces different modulation profiles, and thus different harmonic behaviors.

A detailed statistical study across a larger sample of BHXRBs would be helpful to clarify these source-to-source differences; however, as it falls outside the scope of this work, we do not expand upon it further here. In the following, we focus specifically on the harmonic behavior in MAXI J1348$-$630, which is likely representative of the class including GX 339$-$4 and MAXI J1535$-$571, and explore its underlying physical mechanisms.

\subsection{Intrinsic Nonlinear Distortion in the Hard X-ray Band}
\label{sec:discussion-he}

Bispectral analysis reveals strong phase coupling between the harmonic and the fundamental in the hard X-ray band, even when the harmonic becomes very weak in the late HIMS (see \autoref{fig:pds} and \autoref{fig:rms-spectra}).
This strong phase coupling indicates that the harmonic and the fundamental might share a common physical origin, which can be naturally interpreted as the nonlinear distortion of the QPO waveform. Given that Type-C QPOs are generally attributed to Comptonization processes \citep{Stiele2013, Alabarta_2022, Gao_2023}, this distortion is most likely produced within the corona \citep{Rodriguez_2002_1915, Misra_2013}.

One plausible origin is the geometric projection effect of a precessing coronal region. As the corona precesses, the variation in its projected cross-sectional area is inherently nonlinear, producing harmonic signals alongside the fundamental in the light curve \citep[e.g.,][]{Ingram09, Veledina_2013}. However, since MAXI J1348$-$630 is likely a low-inclination source, this projection effect is expected to be weak. Besides the projection effect, the nonlinear distortion could also arise from the angular beaming of Comptonized emission. In this picture, the angle-dependent flux modulation from the precessing corona induces a nonlinear mapping from the precession phase to the observed flux, giving rise to the harmonic signal \citep{Axelsson14, Axelsson16}. Nevertheless, these purely geometric models have difficulty explaining why the harmonic is significantly detected exclusively during the HIMS.

An alternative explanation is that the nonlinearity arises from the modulation of coronal parameters (e.g., electron temperature and optical depth) within an inhomogeneous corona. Such an inhomogeneous structure can naturally form during the transition from the HS to the HIMS, as enhanced disk emission cools the outer coronal layers and produces a radial gradient \citep{Yamada_2013}. This framework provides a plausible explanation for both the sudden appearance of the significant harmonic at the HIMS onset and its rapid decay. As the source progresses within the HIMS, the corona contracts and becomes more homogeneous, suppressing the nonlinear distortion (and thus the harmonic strength) while preserving the strong phase coupling with the fundamental \citep{Shui_2021, Rawat_2025}.

Notably, any model based on nonlinear waveform distortion faces a key challenge: in the early HIMS, the fractional rms amplitude of the hard X-ray band harmonic is comparable to that of the fundamental, whereas typical nonlinear distortion tends to produce much weaker harmonics \citep{Ingram19, Doesburgh_2020}. This contradiction could be resolved if the fundamental signals from different regions of the inhomogeneous corona are nearly out of phase (\(\Delta\phi\sim\pi\)), causing partial destructive interference that suppresses the fundamental amplitude and thus enhances the relative strength of the harmonic, a physical picture that also aligns with recent Comptonization models \citep{Garc2021}.

\subsection{Disk Reprocessing in the Soft X-ray Band}
\label{sec:discussion-le}

While the hard X-ray harmonic discussed in \autoref{sec:discussion-he} is most likely driven by nonlinear distortion of the fundamental waveform in the hot corona, the soft X-ray harmonic exhibits distinct properties. In particular, it shows no significant phase coupling with the fundamental throughout the HIMS. This behavior suggests that the soft X-ray band harmonic either arises from a physical mechanism completely distinct from that of the fundamental, or originates from a reprocessing mechanism that scrambles the original phase information of the fundamental signal. 

The peak of the harmonic rms spectrum generally falls in the 4--10 keV range, indicating an origin dominated neither by a pure disk component nor by a pure corona. Instead, it points toward a soft Comptonizing component, likely associated with the disk--corona interface \citep{Done_2006, Done_2007, Axelsson14}. Similar behavior has been reported in GX 339$-$4, where the frequency-resolved spectrum of the harmonic is also consistent with a soft Comptonization component \citep{Axelsson16}. One plausible physical interpretation is an illumination scenario, in which the precessing hot corona sweeps across the outer accretion disk \citep[e.g.,][]{Veledina_2015}, producing a double-peaked modulation over one precession cycle that generates a harmonic signal \citep{Doesburgh_2020}.

This framework naturally explains several observed properties of the soft X-ray band harmonic. Under this framework, the generation mechanisms of the fundamental and the harmonic in the soft X-ray band are different, with the former possibly driven by the precession of the relatively cool outer corona and the latter arising from illumination and reflection at the inner disk. Consequently, the harmonic amplitude can potentially become comparable to, or even exceed, that of the fundamental. Furthermore, since the harmonic arises from reprocessing at the disk, the phase coherence between the fundamental and the harmonic is naturally weak or absent \citep{Gierlinski2009, Uttley_14}.

Notably, the disk-corona reprocessing model inherently predicts a soft lag, which contradicts the observed hard lag of the harmonic. This suggests that the phase lag is likely driven by the inward propagation of the precession signal. Specifically, if the precession signal originates in the outer corona and then travels toward the hotter inner regions, it triggers the soft X-ray emission first before reaching the hard X-ray emitting zones, thereby generating the observed hard lag.

\section{Conclusion}
\label{sec:conclusion}
In this work, we present a detailed analysis of the Type-C QPO harmonic in MAXI J1348$-$630 using broadband \textit{Insight}-HXMT observations.
This harmonic exhibits prominent state- and energy-dependent behaviors.
Specifically, the harmonic is significantly detected only during the HIMS.
It shows no significant phase coupling with the fundamental in the soft X-ray band, while clear phase coupling emerges in the hard X-ray band.
Meanwhile, the harmonic maintains a persistently strong rms amplitude in the soft X-ray band, whereas its rms amplitude decays rapidly in the hard X-ray band.
These results indicate that the harmonic has a complex, energy-dependent origin that cannot be fully explained by simple models, such as nonlinear waveform distortion of the fundamental.
We suggest that the hard X-ray band harmonic may arise from nonlinear distortion of the fundamental signal driven by coronal inhomogeneities, while the soft X-ray band harmonic likely originates from a distinct mechanism associated with the illumination of the inner disk.


\section*{acknowledgements}
\label{sec:ack}
This work made use of the publicly available data and software from the \textit{Insight}-HXMT mission. The \textit{Insight}-HXMT project is funded by the China National Space Administration (CNSA) and the Chinese Academy of Sciences (CAS).
X.L.W. \& R.Y.M. were supported by the National Key R\&D Program of China (grant No. 2023YFA1607902);
F.G.X. \& R.Y.M. were supported by the National SKA Program of China (No. 2020SKA0110102);
J.F.W. was supported by the National Key R\&D Program of China (grant No. 2023YFA1607904);
This work was supported in part by the Natural Science Foundation of China (NSFC, grants 12373049, 12361131579, 12373017, 12192220, 12192223, 12033004, 12221003, U2038108 and 12133008).

\section*{Data Availability}
The data underlying this article are available in the \textit{Insight}-HXMT public archive.



\bibliographystyle{aasjournal}
\bibliography{ms} 

\appendix
\section{The fitting results of the harmonic}
\label{appendix:all-fit-value}
\setlength{\LTcapwidth}{\textwidth}
\begin{longtable}{rrccccccc}
    \caption{Best-fitting parameters of the Type-C QPO and its harmonic for exposures where the harmonic detection significance is \(\ge 3\sigma\) in at least one energy band. Parameters without reported errors are tied to the corresponding parameters in other bands, as they cannot be well constrained during fitting. This is particularly the case for the harmonics after MJD 58519, which are less significant in the ME and HE bands (see the lower PDSs in \autoref{fig:pds} for example).
    }
    \label{tab:harmonic_fits} \\

    \toprule
    \multicolumn{1}{c}{\multirow{2}[4]{*}{MJD}} & \multicolumn{1}{c}{\multirow{2}[4]{*}{ExpID}} & \multirow{2}[4]{*}{Energy bands} & \multicolumn{3}{c}{Type-C QPO} & \multicolumn{3}{c}{Harmonic} \\
          &       &       & Freq (Hz) & FWHM (Hz) & Frac rms (\%) & Freq (Hz) & FWHM (Hz) & Frac rms (\%) \\
    \midrule
    \endfirsthead
    \multicolumn{9}{c}{{\bfseries \tablename\ \thetable{} -- continued from previous page}} \\
    \toprule
    \multicolumn{1}{c}{\multirow{2}[4]{*}{MJD}} & \multicolumn{1}{c}{\multirow{2}[4]{*}{ExpID}} & \multirow{2}[4]{*}{Energy bands} & \multicolumn{3}{c}{Type-C QPO} & \multicolumn{3}{c}{Harmonic} \\
          &       &       & Freq (Hz) & FWHM (Hz) & Frac rms (\%) & Freq (Hz) & FWHM (Hz) & Frac rms (\%) \\
    \midrule
    \endhead

    \endfoot

    \bottomrule
    \endlastfoot

    58518.01 & P021400200601 & LE & \(1.03^{+0.10}_{-0.08}\) & \(0.29^{+0.21}_{-0.15}\) & \(4.36^{+1.32}_{-1.12}\) & \(2.42^{+0.06}_{-0.08}\) & \(0.22^{+0.24}_{-0.15}\) & \(3.20^{+1.12}_{-0.73}\) \\
     &  & ME & \(1.12^{+0.01}_{-0.01}\) & \(0.12^{+0.06}_{-0.04}\) & \(4.09^{+0.42}_{-0.44}\) & \(2.35^{+0.05}_{-0.04}\) & \(0.34^{+0.18}_{-0.14}\) & \(3.61^{+0.81}_{-0.59}\) \\
     &  & HE & \(1.09^{+0.02}_{-0.02}\) & \(0.21^{+0.07}_{-0.06}\) & \(4.90^{+0.44}_{-0.47}\) & \(2.34^{+0.04}_{-0.04}\) & \(0.23^{+0.12}_{-0.10}\) & \(3.75^{+0.73}_{-0.73}\) \\
    \midrule
    58518.15 & P021400200602 & LE & \(1.32^{+0.03}_{-0.03}\) & \(0.41^{+0.12}_{-0.10}\) & \(7.01^{+0.68}_{-0.77}\) & \(2.94^{+0.06}_{-0.06}\) & \(0.77^{+0.25}_{-0.19}\) & \(5.74^{+0.87}_{-0.68}\) \\
     &  & ME & \(1.36^{+0.02}_{-0.02}\) & \(0.39^{+0.07}_{-0.06}\) & \(6.30^{+0.45}_{-0.41}\) & \(2.64^{+0.05}_{-0.04}\) & \(0.55^{+0.22}_{-0.17}\) & \(4.59^{+1.03}_{-0.86}\) \\
     &  & HE & \(1.34^{+0.01}_{-0.01}\) & \(0.22^{+0.03}_{-0.03}\) & \(5.03^{+0.34}_{-0.34}\) & \(2.79^{+0.02}_{-0.02}\) & \(0.46^{+0.06}_{-0.06}\) & \(4.64^{+0.38}_{-0.33}\) \\
    \midrule
    58518.29 & P021400200603 & LE & \(1.39^{+0.05}_{-0.07}\) & \(0.25\) & \(4.25^{+0.86}_{-0.93}\) & \(2.93^{+0.10}_{-0.12}\) & \(0.36^{+0.34}_{-0.19}\) & \(3.08^{+1.06}_{-0.69}\) \\
     &  & ME & \(1.48^{+0.02}_{-0.02}\) & \(0.25^{+0.07}_{-0.06}\) & \(5.57^{+0.45}_{-0.45}\) & \(2.90^{+0.07}_{-0.05}\) & \(0.51^{+0.30}_{-0.19}\) & \(4.25^{+1.16}_{-0.91}\) \\
     &  & HE & \(1.44^{+0.01}_{-0.01}\) & \(0.23^{+0.04}_{-0.04}\) & \(6.04^{+0.37}_{-0.39}\) & \(2.91^{+0.04}_{-0.04}\) & \(0.33^{+0.12}_{-0.12}\) & \(3.90^{+0.68}_{-0.60}\) \\
    \midrule
    58519.01 & P021400200701 & LE & \(3.01^{+0.04}_{-0.04}\) & \(0.32^{+0.15}_{-0.11}\) & \(3.69^{+0.65}_{-0.47}\) & \(6.38^{+0.12}_{-0.12}\) & \(1.23^{+0.34}_{-0.31}\) & \(4.18^{+0.48}_{-0.47}\) \\
     &  & ME & \(2.99^{+0.02}_{-0.02}\) & \(0.47^{+0.06}_{-0.06}\) & \(8.27^{+0.43}_{-0.38}\) & \(6.38\) & \(1.23\) & \(1.57^{+1.01}_{-0.59}\) \\
     &  & HE & \(2.97^{+0.07}_{-0.07}\) & \(0.65^{+0.24}_{-0.15}\) & \(9.33^{+0.97}_{-0.87}\) & \(6.38\) & \(1.23\) & \(3.61^{+1.27}_{-1.08}\) \\
    \midrule
    58519.15 & P021400200702 & LE & \(3.66^{+0.07}_{-0.08}\) & \(0.59\) & \(3.62^{+0.38}_{-0.37}\) & \(7.76^{+0.20}_{-0.19}\) & \(1.77^{+0.70}_{-0.54}\) & \(3.33^{+0.59}_{-0.50}\) \\
     &  & ME & \(3.43^{+0.03}_{-0.02}\) & \(0.59^{+0.06}_{-0.07}\) & \(8.81^{+0.45}_{-0.52}\) & \(7.76\) & \(1.77\) & \(1.01^{+1.49}_{-0.43}\) \\
     &  & HE & \(3.50^{+0.01}_{-0.02}\) & \(0.68^{+0.04}_{-0.04}\) & \(10.36^{+0.19}_{-0.20}\) & \(7.76\) & \(1.77\) & \(2.61^{+0.32}_{-0.42}\) \\
    \midrule
    58519.29 & P021400200703 & LE & \(3.76^{+0.13}_{-0.15}\) & \(0.48\) & \(2.14^{+0.29}_{-0.28}\) & \(7.78^{+0.12}_{-0.10}\) & \(1.24^{+0.63}_{-0.45}\) & \(2.60^{+0.42}_{-0.32}\) \\
     &  & ME & \(3.84^{+0.02}_{-0.02}\) & \(0.48^{+0.06}_{-0.05}\) & \(9.35^{+0.38}_{-0.37}\) & \(7.78\) & \(1.24\) & \(1.89^{+1.06}_{-0.66}\) \\
     &  & HE & \(3.71^{+0.02}_{-0.02}\) & \(0.49^{+0.05}_{-0.06}\) & \(10.05^{+0.32}_{-0.42}\) & \(7.78\) & \(1.24\) & \(2.05^{+0.77}_{-0.64}\) \\
    \midrule
    58519.43 & P021400200704 & LE & \(3.92^{+0.06}_{-0.07}\) & \(0.60\) & \(2.99^{+0.29}_{-0.28}\) & \(7.67^{+0.23}_{-0.23}\) & \(3.70^{+1.11}_{-1.03}\) & \(4.23^{+0.82}_{-0.62}\) \\
     &  & ME & \(3.77^{+0.02}_{-0.02}\) & \(0.60^{+0.07}_{-0.06}\) & \(9.76^{+0.46}_{-0.42}\) & \(7.67\) & \(3.70\) & \(3.65^{+1.82}_{-1.35}\) \\
     &  & HE & \(3.71^{+0.01}_{-0.02}\) & \(0.60^{+0.04}_{-0.05}\) & \(10.74^{+0.31}_{-0.34}\) & \(7.67\) & \(3.70\) & \(5.57^{+0.72}_{-0.75}\) \\
    \midrule
    58520.07 & P021400200803 & LE & \(4.42^{+0.24}_{-0.34}\) & \(1.23^{+0.71}_{-0.55}\) & \(2.97^{+0.75}_{-0.58}\) & \(9.08^{+0.26}_{-0.22}\) & \(2.42^{+0.88}_{-0.77}\) & \(3.75^{+0.42}_{-0.38}\) \\
     &  & ME & \(4.20^{+0.02}_{-0.02}\) & \(0.71^{+0.07}_{-0.06}\) & \(10.02^{+0.31}_{-0.27}\) & \(9.08\) & \(2.42\) & \(1.38^{+1.36}_{-0.57}\) \\
     &  & HE & \(4.32^{+0.01}_{-0.02}\) & \(0.51^{+0.04}_{-0.04}\) & \(10.61^{+0.36}_{-0.33}\) & \(9.08\) & \(2.42\) & \(1.96^{+1.07}_{-0.68}\) \\
    \midrule
    58520.21 & P021400200804 & LE & \(4.46^{+0.07}_{-0.07}\) & \(0.37\) & \(2.09^{+0.25}_{-0.24}\) & \(9.28^{+0.15}_{-0.14}\) & \(1.58^{+0.43}_{-0.35}\) & \(2.91^{+0.30}_{-0.26}\) \\
     &  & ME & \(4.52^{+0.02}_{-0.02}\) & \(0.37^{+0.07}_{-0.06}\) & \(8.03^{+0.57}_{-0.56}\) & \(9.28\) & \(1.58\) & \(1.66^{+1.27}_{-0.70}\) \\
     &  & HE & \(4.52^{+0.02}_{-0.01}\) & \(0.51^{+0.05}_{-0.05}\) & \(10.52^{+0.41}_{-0.44}\) & \(9.28\) & \(1.58\) & \(1.63^{+0.95}_{-0.62}\) \\
    \midrule
    58520.60 & P021400200901 & LE & \(4.91^{+0.06}_{-0.05}\) & \(0.61\) & \(2.21^{+0.18}_{-0.20}\) & \(9.91^{+0.18}_{-0.14}\) & \(1.48^{+0.51}_{-0.36}\) & \(2.33^{+0.32}_{-0.24}\) \\
     &  & ME & \(5.00^{+0.01}_{-0.01}\) & \(0.65^{+0.05}_{-0.05}\) & \(10.83^{+0.30}_{-0.33}\) & \(9.91\) & \(1.48\) & \(1.36^{+1.17}_{-0.51}\) \\
     &  & HE & \(4.99^{+0.02}_{-0.02}\) & \(0.61^{+0.07}_{-0.07}\) & \(10.54^{+0.64}_{-0.55}\) & \(9.91\) & \(1.48\) & \(1.14^{+1.07}_{-0.49}\) \\
    \midrule
    58521.46 & P021400201001 & LE & \(5.53^{+0.16}_{-0.17}\) & \(1.62^{+0.47}_{-0.43}\) & \(2.69^{+0.32}_{-0.40}\) & \(11.02^{+0.26}_{-0.34}\) & \(3.76^{+1.13}_{-0.90}\) & \(2.82^{+0.32}_{-0.25}\) \\
     &  & ME & \(5.38^{+0.03}_{-0.03}\) & \(0.74^{+0.10}_{-0.09}\) & \(10.33^{+0.87}_{-0.77}\) & \(11.02\) & \(3.76\) & \(3.39^{+1.28}_{-1.12}\) \\
     &  & HE & \(5.35^{+0.03}_{-0.02}\) & \(0.65^{+0.10}_{-0.09}\) & \(9.37^{+0.75}_{-0.60}\) & \(11.02\) & \(3.76\) & \(1.97^{+1.59}_{-0.82}\) \\
    \midrule
    58521.76 & P021400201003 & LE & \(5.24^{+0.11}_{-0.11}\) & \(0.94^{+0.40}_{-0.30}\) & \(2.58^{+0.38}_{-0.36}\) & \(11.06^{+0.29}_{-0.29}\) & \(3.32^{+1.23}_{-0.84}\) & \(3.31^{+0.34}_{-0.31}\) \\
     &  & ME & \(5.30^{+0.02}_{-0.02}\) & \(0.54^{+0.06}_{-0.06}\) & \(9.58^{+0.37}_{-0.39}\) & \(11.06\) & \(3.32\) & \(2.45^{+1.38}_{-0.99}\) \\
     &  & HE & \(5.30^{+0.02}_{-0.02}\) & \(0.64^{+0.06}_{-0.07}\) & \(10.84^{+0.52}_{-0.59}\) & \(11.06\) & \(3.32\) & \(1.85^{+1.51}_{-0.75}\) \\
    \midrule
    58600.74 & P021400205301 & LE & \(7.04^{+0.21}_{-0.21}\) & \(1.43^{+0.83}_{-0.63}\) & \(3.95^{+0.77}_{-0.66}\) & \(14.82^{+0.29}_{-0.27}\) & \(1.76^{+1.11}_{-0.69}\) & \(3.65^{+0.56}_{-0.55}\) \\
     &  & ME & \(7.08^{+0.17}_{-0.14}\) & \(1.67^{+2.05}_{-1.05}\) & \(6.48^{+2.73}_{-1.92}\) & \(14.82\) & \(1.76\) & \(3.76^{+3.77}_{-1.58}\) \\
     &  & HE & \(7.14^{+0.13}_{-0.20}\) & \(0.32^{+0.47}_{-0.24}\) & \(6.62^{+4.03}_{-2.26}\) & \(14.82\) & \(1.76\) & \(7.45^{+2.64}_{-1.96}\) \\
    \midrule
    58602.09 & P021400206902 & LE & \(4.33^{+0.08}_{-0.08}\) & \(0.61^{+0.08}_{-0.03}\) & \(4.44^{+0.48}_{-0.42}\) & \(8.58^{+0.36}_{-0.38}\) & \(1.80^{+0.85}_{-1.26}\) & \(3.40^{+0.79}_{-0.56}\) \\
     &  & ME & \(4.33\) & \(0.61\) & \(4.64^{+1.39}_{-1.87}\) & \(8.58\) & \(1.80\) & \(4.17^{+1.54}_{-2.87}\) \\
     &  & HE & \(4.33\) & \(0.61\) & \(5.20^{+1.20}_{-1.45}\) & \(8.58\) & \(1.80\) & \(3.61^{+1.38}_{-2.27}\) \\
    \midrule
    58602.22 & P021400206903 & LE & \(4.21^{+0.16}_{-0.15}\) & \(1.09^{+0.40}_{-0.75}\) & \(4.87^{+0.86}_{-1.35}\) & \(8.39^{+0.17}_{-0.17}\) & \(0.72^{+0.32}_{-0.49}\) & \(3.06^{+0.59}_{-0.52}\) \\
     &  & ME & \(4.21\) & \(1.09\) & \(4.53^{+1.52}_{-1.44}\) & \(8.39\) & \(0.72\) & \(4.09^{+1.65}_{-2.55}\) \\
     &  & HE & \(4.21\) & \(1.09\) & \(5.80^{+1.25}_{-1.34}\) & \(8.39\) & \(0.72\) & \(5.46^{+1.65}_{-1.81}\) \\
\end{longtable}
\label{lastpage}
\end{document}